\documentstyle[12pt]{article}
\newcommand{\Frac}[2]{\frac{\displaystyle #1}{\displaystyle #2}}
\begin{document}
\pagestyle{empty}
\input{feynman}
\begin{titlepage}
\begin{center}
\vspace*{-2cm}
\hfill FTUV/94-24\\
\hfill IFIC/94-22\\
\hfill DTP/94-28 \\
\vspace{1cm}
{\Large \bf Parameter free calculation of non--resonant \\
\vspace*{0.3cm}
three body decays of charmed mesons
\vspace*{0.3cm} \\ in a
weak gauged  \vspace*{0.4cm} \\   $U(4)_L \otimes
U(4)_R$ Chiral Lagrangian Model} \\
\vspace{1cm}
{\large F.J. Botella, S. Noguera} \\
\vspace*{0.3cm}
Departament de F\'{\i}sica Te\`orica and I.F.I.C. (Centre Mixte
Universitat de Val\`encia--C.S.I.C.) \\
E-46100 Burjassot (Val\`encia, Spain) \\
\vspace*{0.7cm}
{\large J. Portol\'es} \footnote{Now at Particle Theory Division in
Rutherford--Appleton Laboratory, Chilton, DIDCOT, OX11 0QX (U.K.)} \\
\vspace*{0.3cm}
Physics Department, University of Durham \\
Durham , DH1 3LE (U.K.)
\vspace*{0.5cm} \\
\begin{abstract}
Non--resonant decays of charmed mesons into three pseudoscalars are
analyzed in a weak gauged $U(4)_L \otimes U(4)_R$ chiral lagrangian
model. The calculation is free of unknown parameters and only requires
 the masses of pseudoscalar mesons as hadronic inputs. When comparison
with experimental data is possible we find that in some processes we
have good agreement and in others we are an order of magnitude below.
This may be due to  the absence in our calculation
of final state interactions, presumably important here, and the
manifest difficulty in extracting the non--resonant contribution
from an experimental point of view.
\end{abstract}
\end{center}
\end{titlepage}
\newpage
\pagestyle{plain}
\pagenumbering{arabic}
\vspace{4cm}

\hspace{0.5cm}Non--leptonic weak decays of charmed mesons have received
in the last few years  increasing attention both experimentally
and theoretically. The amount of data, now available
\cite{PDG}, has motivated  many authors to study  theoretical
prescriptions for these processes in a framework in which non--leptonic weak
decays have always been a challenge for the standard theory. The point
usually employed is to take the effective four quark operators that
allow the weak process and try to impose the strong effects of QCD on these.
 However,
non--perturbative effects, which are important in hadrons, introduce
some freedom in the choice of assumptions needed to calculate
hadron matrix elements as has been manifested in the widely studied
two body processes \cite{ch2,ch3,ch4,ch5}. Besides non--leptonic
decays of charmed mesons are hard to predict because of the
rescattering corrections required by unitarity and the fact that
the matrix elements are evaluated in an energy range populated by
many strong resonances \cite{ch6}. Therefore the decays of charmed mesons
into three pseudoscalars are, in general,
dominated by vector resonances: the quasi--two--body decays containing
a vector meson (mainly $\rho$ or $K^*$) and a pseudoscalar ($K$ or $\pi$).
Consequently, non--resonant contributions are usually only small
fractions of the total $D \rightarrow P P P$ decay. Experimentally
it seems hard to measure the direct three body decays because of the
possible background effects and the interference between non--resonant
and quasi--two--body amplitudes.
\par
The direct three body decays have not received too much attention from
a theoretical point of view. A general outline has been obtained by using
a bosonization of the hadron currents in a chiral lagrangian \cite{cheng},
the quark diagram approach \cite{chau} and, tentatively, the factorization
procedure \cite{kapipi}.
\par
In this paper we present the analysis of decays of charmed mesons
into three pseudoscalars ($D \rightarrow P P P$) in a weak gauged
$U(4)_L \otimes U(4)_R$ chiral lagrangian model we have presented
in reference \cite{we}. In the same paper, the model has been tested
at one loop level in radiative rare kaon decays ($K^+ \rightarrow
\pi^+ l^+ l^-$ , $K_S \rightarrow \pi^{\circ} l^+ l^-$) with satisfactory
results. It has also been applied to $D \rightarrow P P$ processes in
\cite{charm2} giving a reasonable pattern for the calculated widths.
In this last paper several predictions were given.
\par
Recently, the CLEO Collaboration has measured the branching ratio
$B(D^{\circ} \rightarrow K^+ \pi^-) / B(D^{\circ} \rightarrow
K^- \pi^+) = 0.0077 \pm 0.0025 (stat) \pm 0.0025 (syst)$ \cite{cinabro}.
Using the known value of $B(D^{\circ} \rightarrow K^- \pi^+)$ \cite{PDG}
we obtain for $\Gamma (D^{\circ} \rightarrow K^+ \pi^-)$
the value $(4.4 \pm 1.4) \times 10^{-16} GeV$, where only
the statistical error has been considered. This result is in good
agreement with our prediction in \cite{charm2} $\Gamma (D^{\circ}
\rightarrow K^+ \pi^-) = 3.3 \times 10^{-16} GeV$.

\par
The basic idea of our model is to solve the strong interaction
by using a chiral lagrangian and then introduce the weak
interactions through the requirement of local gauge invariance under
the electroweak group $SU(2)_L \otimes U(1)_Y$ over the
 mesonic degrees of freedom. In order to have two
doublets under $SU(2)_L$ we are forced to enlarge the chiral lagrangian,
so we have chosen a $U(4)_L \otimes U(4)_R$ chiral lagrangian. In
particular we have taken a linear realization in order to preserve
predictability as much as possible. This model  naturally includes not
only all the symmetries of the two--generation Standard Theory but
also its symmetry breaking patterns.
\vspace*{0.5cm} \\
\hspace{0.5cm}We refer the reader to the references \cite{we,charm2} for a
detailed explanation of our model. We give here a brief outline.
\par
Our lagrangian is:
\begin{equation}
       {\cal L} = {\cal L}_{meson} + {\cal L}_{Higgs}
                  + {\cal L}_{HM}  + \ldots
\end{equation}
where the dots are short for pure gauge boson terms.
Here ${\cal L}_{meson}$ contains the strong interaction between mesons
and the couplings of mesons to the gauge bosons, ${\cal L}_{Higgs}$
is the usual lagrangian for the minimal model of Higgs of the Standard
Theory and ${\cal L}_{HM}$ is a Higgs--meson coupling term which will give
masses to the mesons after the spontaneous breaking of the weak symmetry
(it has been detailed in \cite{we,charm2}).
We have a set of $16$ scalars and $16$ pseudoscalars fields which
we assign to the
$(4,\bar{4})$ representation of the chiral
$U(4)_L \otimes U(4)_R$ group. We denote the meson matrix by
$U = \Sigma + i \Pi$, where $\Sigma$ is the scalar and $\Pi$
the pseudoscalar matrices of fields.
The explicit expression for the pseudoscalar matrix is
\begin{eqnarray*}
      \Pi & = & \left( \begin{array}{cccc}
                   \frac{1}{2}\eta_{\circ} +
                   \frac{1}{\sqrt{2}}\pi^{\circ} + & &
                   &  \\ & \pi^{+} & K^{+} & \overline{D^{\circ}} \\
                   \frac{1}{\sqrt{6}}\eta_8 + \frac{1}{\sqrt{12}}
                   \eta_{15} & & & \\
                   & & & \\ & & & \\
                    & \frac{1}{2}\eta_{\circ} -
                   \frac{1}{\sqrt{2}}\pi^{\circ} + & & \\
                   \pi^{-} & & K^{\circ} & D^{-} \\
                   & \frac{1}{\sqrt{6}}\eta_8 +
                   \frac{1}{\sqrt{12}}\eta_{15} & & \\
                   & & & \\ & & & \\
                   & & \frac{1}{2}\eta_{\circ} -
                   \frac{2}{\sqrt{6}}\eta_8 + & \\
                   K^{-} & \overline{K^{\circ}} & & D_S ^{-} \\
                   & & \frac{1}{\sqrt{12}}\eta_{15} & \\
                   & & & \\ & & & \\
                   D^{\circ} & D^{+} & D_S ^{+} &
                   \frac{1}{2}\eta_{\circ} -
                   \frac{3}{\sqrt{12}}\eta_{15}
                   \end{array}
             \right)
\end{eqnarray*}
\begin{equation}
         \;
\end{equation}

A similar matrix can be written for the scalar mesons. Our notation
for scalar mesons is $\sigma_{\circ},\sigma_8, \sigma_{15}, \sigma^+,
\sigma_3, \kappa, \delta, \delta_S$ instead of $\eta_{\circ},
\eta_8,\eta_{15}, \pi^+, \pi^{\circ}, K, D $ and $D_S$ respectively.
\par
With these definitions ${\cal L}_{mesons}$ is

\begin{equation}
      {\cal L}_{mesons} = \frac{1}{2} Tr[(D^{\mu}U')^{\dagger}
                                         (D_{\mu}U')] -
                           V_{chiral}(U)
\end{equation}
where $V_{chiral}$ is the chiral potential
\begin{eqnarray}
      V_{chiral}(U) & = & - \mu_{\circ}^2 \; Tr(U^{\dagger}U) +
                            \nonumber \\*
                    &   &
                            \mu_{\circ}^2 \; [  a \;
                                 Tr(U^{\dagger}U)^2  +
                                        b \; (Tr(U^{\dagger}U))^2  +
                                        c \; (det U + det U^{\dagger}
                                            )  ]
\end{eqnarray}
with $\mu_{\circ}^2 > 0$ in order to develop spontaneous breaking
of chiral symmetry.
The covariant derivative is:
\begin{equation}
      D_{\mu} U' = \partial_{\mu} U' -
                        i g \vec{T} \cdot \vec{W_{\mu}} U' -
                        i g' Y_L B_{\mu} U' +
                        i g' U' Y_R B_{\mu}
\end{equation}
with
\begin{equation}
              U' = S U S^{\dagger}
\end{equation}
Here $\vec{W_{\mu}}$ and $B_{\mu}$ are the gauge bosons related to
the $SU(2)_L$ and the $U(1)_Y$ groups. The $\vec{T}$ matrices are the
$SU(2)$ generators and $Y_L$ and $Y_R$ are the left and right
hypercharges. The matrix $S$ will be the Cabibbo rotation once
the $SU(2)_L \otimes U(1)_Y$ symmetry gets spontaneously  broken.
With our definitions for $U$ we have
\begin{equation}
\begin{array}{cc}
      \vec{T} = \Frac{1}{2} \left( \begin{array}{cc}
                                   \vec{\tau} & 0 \\
                                   0 & \tau^1 \vec{\tau} \tau^1
                                   \end{array} \right)
      &
      Y_L = \Frac{1}{6} I_{4\times4} \\    \\
      Y_R = \Frac{1}{3} \left( \begin{array}{cccc}
                               2 &  &  &  \\
                                 & -1 &  &  \\
                                 &  & -1 &  \\
                                 &  &  & 2
                               \end{array} \right)
      & \, \; \, \;
      S = \left( \begin{array}{cccc}
                 1 & 0  & 0  & 0 \\
                 0 & \cos \theta_c & \sin \theta_c & 0 \\
                 0 & - \sin \theta_c & \cos \theta_c & 0 \\
                 0 & 0 & 0 & 1
                 \end{array} \right)
\end{array}
\end{equation}
with $\vec{\tau}$ the usual Pauli matrices and $\theta_c$ the
Cabibbo angle. The charge operator is $Q = Y_R = Y_L + T_3$.
\par
Through the spontaneous breaking of the chiral and weak
symmetry the matrices of mesons and Higgs get a non--zero vacuum
expectation value
\begin{equation}
\begin{array}{cc}
\langle \circ | U | \circ \rangle \equiv F = \Frac{1}{\sqrt{2}}
\left( \begin{array}{cccc}
        f_{\alpha} & & & \\
        & f_{\alpha} & & \\
        & & f_{\gamma} & \\
        & & & f_{\delta}
        \end{array} \right) , \; \; \;  &
\langle \circ | H | \circ \rangle \equiv  \Frac{1}{\sqrt{2}} \phi_{\circ}
I_{4 \times 4}
\end{array}
\label{eq:ff}
\end{equation}
where we have assumed exact isospin symmetry.
\par
We have been working in a limit in which the
masses of the scalars are very big in comparison with those of the
pseudoscalars. This
fact is equivalent to taking the $\mu_{\circ}^2 \rightarrow \infty$
limit but keeping $\mu_{\circ}^2 c \equiv c'$ constant in order
to maintain the $\eta_{\circ}$ mass finite \cite{Witten}. In this case
$F$ in (\ref{eq:ff}) becomes
\begin{equation}
F = \Frac{1}{\sqrt{2}} f_{\circ} I_{4 \times 4}
\end{equation}
\par
As we have said above we will assume $SU(2)_I$ symmetry in our calculation.
 At this level the only hadronic inputs will be the masses
$m_{\pi}$, $m_K$ and $m_D$. In the large $\mu_{\circ}^2$ limit,
once those values have been chosen the only free parameter of our
lagrangian is $c'$. The value for $c'$ is fixed from $m_{\eta},
m_{\eta'}$ and $m_{\eta_c}$. These three masses have a weak dependence
on $c'$. A
reasonable range of variation for $c'$ is $c' = -33, -28$. For
these values of $c'$ the masses are $m_{\eta}= \, 0.507 \, GeV, \,
0.497 \, GeV$, $m_{\eta'} = \, 1.023 \, GeV, \, 0.969 \, GeV$ and
$m_{\eta_c} = \, 2.701 \, GeV, \, 2.691 \, GeV$ respectively.
\par
As a first approach to the processes $D \rightarrow PPP$ we will
simply present a tree level calculation. Therefore we do not give
predictions for $D^{\circ} \rightarrow \bar{K^{\circ}} \pi^{\circ}
\pi^{\circ}$, $K_S^{\circ} K_S^{\circ} K_S^{\circ}$, ... The goal
of our paper is to give a general pattern for the considered
processes by using a procedure where no free parameters are left.
Then we have not taken into account final--state strong interactions
presumably  important here as we have said above.
Such calculation, however, would not be well defined in a chiral
model with heavy pseudoscalar mesons. Starting with an $U(4)_L \otimes
 U(4)_R$ chiral lagrangian in order to be able to implement the
$SU(2)_L \otimes U(1)$ weak symmetry \cite{we}, we have introduced an
excess of symmetry in the strong sector (chiral symmetry is
badly broken due to the large masses of the charmed pseudoscalars) .
 The next to leading strong
interaction corrections would be sensitive to that excess of
symmetry. In any case it is not our goal
to discuss the strong interaction sector of the model but to study
the weak transitions
involved in the weak decays of charmed mesons.
\par
In order to keep straightforward predictability we have chosen the
limit reported above in which the scalar masses are very large in
comparison with the pseudoscalar masses (limit $\mu_{\circ}^2 \rightarrow
\infty$, $\mu_{\circ}^2 c \equiv c'$). This is in fact the only
sensible way to calculate in the actual stage of the model without
including unknown parameters. The scalar
sector of our model only can be reliably taken when the whole
hadronic spectrum in the energy range of interest (vectors,
axial--vectors, ...) is properly included \cite{we,Ko}. Therefore
the introduction of the scalar mesons as the model rules would input
uncontrolled information. The r\^ole of the scalar mesons is that of
introducing a higher energy scale than $\Lambda_{\chi} \sim 4 \pi
f_{\pi}$ that gives meaning to the chiral expansion once the
explicit chiral symmetry breaking scale ($m_D$) is included.
Nevertheless, we have studied some processes by taking finite
reasonable masses (of a few GeV) for the scalar mesons, without
changes in the final results.
\par
There are two different kinds of diagrams which can contribute to the
general process $D \rightarrow P_1 P_2 P_3$ as shown in Fig. 1: those
with an {\em external} weak transition (Fig. 1.a)  (it can happen in
one of the four external legs \footnote{Shortly we distinguish between
final external and initial external transition depending on where the weak
vertex is placed.} and
therefore we can factorize the strong and weak sectors) and
those where there is an {\em inner} weak transition (Fig. 1.b). To perform
the calculations we have chosen a renormalizable $R_{\xi}$ gauge
\cite{Fuji} whose peculiar realization has been discussed in \cite{we}.
In this gauge the usual direct coupling between mesons and $W_{\mu}$
is replaced by a mixing between mesons and $s^{+}$, the charged
Higgs.
\par
The amplitudes for the processes can be written in the following
form:
\begin{itemize}
\item[a)] Cabibbo allowed.
\begin{equation}
A_{cc}(D \rightarrow P_1 P_2 P_3) = G_F \cos^2 \theta_c
\tilde{A}_{cc}(D \rightarrow P_1 P_2 P_3)
\end{equation}
\item[b)] Cabibbo suppressed.
\begin{equation}
A_{cs}(D \rightarrow P_1 P_2 P_3) = G_F \sin \theta_c \cos \theta_c
\tilde{A}_{cs}(D \rightarrow P_1 P_2 P_3)
\end{equation}
\item[c)] Double Cabibbo suppressed.
\begin{equation}
A_{ss}(D \rightarrow P_1 P_2 P_3) = G_F \sin^2 \theta_c
\tilde{A}_{ss}(D \rightarrow P_1 P_2 P_3)
\end{equation}
\end{itemize}
where the $\tilde{A}(D \rightarrow P_1 P_2 P_3)$ factor splits up
(in the general case) into three possible contributions
$\tilde{A} = A_{ext}^{ann} + A_{ext} + A_{inn}$. $A_{ext}^{ann}$
comes from diagrams in Fig. 1.a where the external weak transition
happens in the initial $D$ state (only for $D^{\pm}$ and $D_S^{\pm}$),
it is a kind of {\em annihilation--like} diagram. $A_{ext}$ comes from
diagrams in Fig. 1.a where the external weak transition happens
in one of the electrically charged final states. In $A_{ext}^{ann}$ and
$A_{ext}$ we can factorize each amplitude in a strong and a weak
factors. Finally, $A_{inn}$
represents the inner weak transition in Fig. 1.b that is not factorizable.
\par
$A_{ext}^{ann}$ is suppressed over $A_{ext}$ by a factor  of
$m_{K,\pi}^2 / m_D^2$ typically. This is because
in the factorization the weak factor (Fig. 2) has a different behaviour
depending where it occurs:
the $\Pi_1$ state in Fig. 2.a
has no charm and so corresponds to $K^{\pm}$ or $\pi^{\pm}$ while $\Pi_2$ in
Fig. 2.b has still the charm degree of freedom and therefore it is
a $D^{\pm}$ or $D_S^{\pm}$.
\par
Let us illustrate these facts with a detailed explanation of two processes:
$D^+ \rightarrow \pi^+ \pi^+ \pi^-$ and $D^{\circ} \rightarrow
K^{\circ} K^- \pi^+$.
\vspace*{0.75cm} \\
\underline{{\bf $D^+ \rightarrow \pi^+ \pi^+ \pi^-$}}
\vspace*{0.5cm} \\
Only factorizable diagrams like those in Fig. 1.a contribute to this
process. The two external weak transitions can be factorized as
\begin{equation}
A_{ext}^{ann} (D^+ \rightarrow \pi^+ \pi^+ \pi^-) =
A_{weak}^{ann}(D^+ \rightarrow \pi^+) \times
A_{strong}^{ann} (\pi^+ \rightarrow \pi^+ \pi^+ \pi^-)
\label{eq:nn1}
\end{equation}
and
\begin{equation}
A_{ext}(D^+ \rightarrow \pi^+ \pi^+ \pi^-) =
A_{strong} (D^+ \rightarrow D^+ \pi^+ \pi^-) \times
A_{weak}(D^+ \rightarrow \pi^+)
\label{eq:nn2}
\end{equation}
where $A_{weak}^{ann}(D^+ \rightarrow \pi^+)$, corresponding to Fig. 2.a,
is given by
\begin{equation}
A_{weak}^{ann} (D^+ \rightarrow \pi^+) =
- \sqrt{2} G_F \sin \theta_c \cos \theta_c f_{\circ}^2 \Frac{m_{\pi}^2}
{m_D^2 - m_{\pi}^2}
\label{eq:annw}
\end{equation}
while $A_{weak}(D^+ \rightarrow \pi^+)$, from Fig. 2.b, is
\begin{equation}
A_{weak}(D^+ \rightarrow \pi^+) =
\sqrt{2} G_F \sin \theta_c \cos \theta_c f_{\circ}^2
\Frac{m_D^2}{m_D^2 - m_{\pi}^2}
\label{eq:wea}
\end{equation}
The strong amplitudes in (\ref{eq:nn1}) and (\ref{eq:nn2}) are
\begin{eqnarray}
A_{strong}^{ann} ( \pi^+ (k) \rightarrow \pi^+(p_1) \pi^+(p_2) \pi^-) & = &
  \Frac{i}{f_{\circ}^2} \, [ \, (k-p_1)^2 + (k-p_2)^2 - 2 m_{\pi}^2  \, ]
\nonumber \\
& & \\
A_{strong}(D^+(k) \rightarrow D^+(p_1) \pi^+(p_2) \pi^-) & = &
\Frac{i}{f_{\circ}^2} \, [ \, (k-p_1)^2 + (k-p_2)^2 - m_D^2 - m_{\pi}^2 \, ]
\nonumber
\end{eqnarray}
As can be seen by comparing (\ref{eq:annw}) and (\ref{eq:wea}) the
contribution coming from $A_{ext}^{ann}$ is suppressed over the one
coming from $A_{ext}$ a factor of $m_{\pi}^2 / m_{D}^2$. The origin of
this suppression comes from the weak sector of the amplitudes as explained
above.
\vspace*{1cm} \\
\underline{{\bf $D^{\circ} \rightarrow K^{\circ} K^- \pi^+$}}
\vspace*{0.5cm} \\
As we do not have flavour changing neutral currents at tree level it is
not possible to have an annihilation diagram for this process. Therefore the
non vanishing amplitudes are $A_{ext}$  and $A_{inn}$.
\par
For $A_{ext}$ we have diagrams like those in Fig. 1.a which we factorize
as
\begin{equation}
A_{ext} (D^{\circ} \rightarrow K^{\circ} K^- \pi^+) =
A_{strong}(D^{\circ} \rightarrow K^{\circ} K^- D^+) \times
A_{weak}(D^+ \rightarrow \pi^+)
\end{equation}
where
\begin{eqnarray}
A_{strong}(D^{\circ}(k) \rightarrow K^{\circ} K^-(p_1) D^+(p_2)) & = &
\Frac{i}{2 f_{\circ}^2} \, [  (k-p_1)^2 + (k-p_2)^2 - m_D^2 - m_K^2 \, ]
\nonumber \\
& & \\
A_{weak}(D^+ \rightarrow \pi^+) & = & \sqrt{2} G_F \sin \theta_c
\cos \theta_c f_{\circ}^2 \Frac{m_D^2}{m_D^2 - m_{\pi}^2} \nonumber
\end{eqnarray}
For the internal weak amplitude $A_{inn}$ we have the contribution
of the diagrams in Fig. 1.b
\begin{equation}
A_{inn} (D^{\circ}(k) \rightarrow K^{\circ}(p_1) K^- \pi^+(p_2)) =
-i \Frac{G_F}{\sqrt{2}} \sin \theta_c \cos \theta_c \, [ \, (k-p_1)^2 -
(k-p_2)^2 \, ]
\end{equation}
\vspace*{1cm} \\
By looking to the different amplitudes for each process we see that in
$D_S^+ \rightarrow \pi^+ \pi^+ \pi^-$ and $D^+ \rightarrow K^+ K^+ K^-$
the only contribution comes from $A_{ext}^{ann}$. This fact is easy to
understand because these processes have also only this kind of contribution
in the quark model: they go through an annihilation diagram.
\par
To
$D^{\circ} \rightarrow \overline{K^{\circ}} \pi^+ \pi^-$,
$D^+ \rightarrow K^- \pi^+ \pi^+$ and $D^{\circ} \rightarrow K^{\circ}
K^+ K^-$ only $A_{ext}$ contributes. A non vanishing  $A_{ext}^{ann}$
is forbidden in $D^+ \rightarrow K^- \pi^+ \pi^+$ due to the fact
that by annihilating $D^+$ we can get a $K^+$ but not a $K^-$.
Finally $D^{\circ} \rightarrow K^+ K^- \overline{K^{\circ}}$ and
$D_S^+ \rightarrow \pi^+ \pi^{\circ} \eta \, , \, \pi^+ \pi^{\circ} \eta'$
only get contribution
from $A_{inn}$.
\par
For the rest of processes the decays of $D^{\circ}$ and $D^+ \rightarrow
\overline{K^{\circ}} \pi^+ \pi^{\circ}$ have contributions from $A_{ext}$ and
$A_{inn}$, and the remaining decays of $D^{\pm}$ and $D_S^{\pm}$ from
$A_{ext}^{ann}$ and $A_{ext}$.

In order to calculate the decay widths we have to integrate the
squared amplitude  over all phase space. All the phase space integrals
can be reduced to a one dimensional integral which we calculate numerically.
The numerical results for the processes are given in Tables 1--3
where they are compared with experimental data \cite{PDG,WA82}
where possible.
\par
As can be seen the experimental situation is quite poor. Only a few processes
have been measured and in these the errors are big. In the
comparison some comments are in order:
\begin{itemize}
\item[a)] We have good agreement in $D^{\circ} \rightarrow K^- \pi^+
\pi^{\circ}$, $D^+ \rightarrow \overline{K}^{\circ} \pi^+ \pi^{\circ}$,
$D_S^+ \rightarrow \pi^+ \pi^{\circ} \eta$, $D_S^+ \rightarrow \pi^+
\pi^{\circ} \eta' $, $D^{\circ} \rightarrow K^{\circ} K^- \pi^+$,
$D^+ \rightarrow \pi^+ \pi^+ \pi^-$, within the experimental errors.
\item[b)]There are two processes: $D_S^+ \rightarrow \pi^+ \pi^+ \pi^-$
and $D^+ \rightarrow K^+ K^+ K^-$ where our prediction is orders of magnitude
too small. They deserve separate comments. We understand our result for
$D_S^+ \rightarrow \pi^+ \pi^+ \pi^-$. This process goes through an
annihilation diagram in the quark model (at leading order). In our
calculation the only contribution to this process comes from $A_{ext}^{ann}$
and therefore it is very suppressed. It should be improved in a
next--to--leading calculation. $D^+ \rightarrow K^+ K^+ K^-$ has an
experimental width that is hard to understand (even considering its
big error). This is a double Cabibbo suppressed process that is also
delayed by the small available phase space and in spite of that the
experimental width has the same order of magnitude as, for example,
$D^+ \rightarrow \pi^+ \pi^+ \pi^-$ that is only once Cabibbo suppressed.
\item[c)] The other five
measured processes are one order of magnitude above our predictions. This is
likely to be because our calculations take no account of final state
interactions. Moreover the experimental status is very poor. Indeed in
$D^+ \rightarrow K^- \pi^+ \pi^+$ the measured non--resonant width comes
from an
unexpected non--uniform contribution in the Dalitz plot
\cite{kapipi,MARKIII}.
The extraction of resonant contribution in $D \rightarrow P P P$,
the background and possible interference between resonant and
non--resonant contributions make a precise evaluation of the
direct three body processes difficult from an experimental point of view.
\item[d)]Our theoretical results show, in general, the pattern of
suppression coming from
the phase space available for each process. In contrast the experimental
results do not show this pattern, as can be seen by comparing the
rate of the following processes: $D_S^+ \rightarrow \pi^+ \pi^+ \pi^-$,
$D^+ \rightarrow K^- \pi^+ \pi^+$, $D_S^+ \rightarrow K^+ K^- \pi^+$
in Table I , or $D^+ \rightarrow \pi^+ \pi^+ \pi^-$, $D^{\circ}
\rightarrow \overline{K}^{\circ} K^+ \pi^-$ in Table II.
\end{itemize}
It is worthwhile to emphasize that our calculation is free of unknown
parameters: the only hadronic inputs are just meson masses.
The model we have employed has the symmetries of the Standard Theory,
therefore, at tree level, we have the stricture of a G.I.M. mechanism.
Better results could be obtained by including final state interactions
where that restriction does not operate for neutral particles.
Nevertheless, the large mass of the charmed pseudoscalar mesons prevents
us from using our model to calculate the next to leading
 strong interaction corrections. More work in the understanding of the
model and in order to improve it is needed.
\par
We have seen that a reasonable chiral lagrangian of mesons implementing
 the weak interactions by requiring  local gauge invariance
under the electroweak group allows us to study the non--leptonic decays
of charmed mesons in a parameter-free fashion using a simple calculational
scheme to give an acceptable
pattern of $D \rightarrow P P$ \cite{charm2} and $D \rightarrow P P P$ decays
within their large experimental uncertainties.

\section*{Acknowledgements}
\hspace{0.5cm}The authors are indebted to J. Bernab\'eu for stimulating
this study and to M.R. Pennington for a careful reading of the manuscript.
This work has been supported
by the CICYT under grant AEN93-0234, by DGICYT under grant
PB94-0080 and also, in part, by I.V.E.I.
\par
J.P. has been supported in part by EC Human and Capital Mobility
programme EuroDa$\phi$ne network under grant ERBCMRXCT 920026.

\newpage
\hspace*{-0.6cm}{\Large \bf Table Captions}
\\
\\
\\
{\bf Table I} Comparison of theoretical and experimental widths
 for Cabibbo allowed decays. a) Value predicted for $c' = -33$.
b) Value predicted for $c' = -28$.
\\
\\
{\bf Table II} Comparison of theoretical and experimental widths
 for Cabibbo suppressed decays. a) and b) as in Table I.
\\
\\
{\bf Table III} Comparison of theoretical and experimental widths for double
Cabibbo suppressed decays.
\\
\newpage

\newpage
\begin{center}
{\bf Table I}
\end{center}
$$
\begin{array}{|c|c|c|}
\hline
\hline
\multicolumn{1}{|c}{Decay} &
\multicolumn{1}{|c}{\Gamma_{exp} \times 10^{14} \; \, (GeV)} &
\multicolumn{1}{|c|}{\Gamma_{theor} \times 10^{14} \; \, (GeV)} \\
\hline
\hline
& & \\
D^{\circ} \rightarrow K^- \pi^+ \pi^{\circ} &  1.0 \pm 0.4 & 1.2  \\
& & \\
\hline
& & \\
D^{\circ} \rightarrow \overline{K}^{\circ} \pi^+ \pi^- & 2.3 \pm 0.4 & 0.3  \\
& & \\
\hline
& & \\
D^{\circ} \rightarrow K^+ K^- \overline{K}^{\circ} & 0.8 \pm 0.1  &  0.01  \\
& (non-\phi) & \\
\hline
& & \\
D^+ \rightarrow K^- \pi^+ \pi^+  & 4.5 \pm 0.9 &  0.6 \\
& & \\
\hline
& & \\
D^+ \rightarrow \overline{K}^{\circ} \pi^+ \pi^{\circ}  & 0.8 \pm 0.7 & 1.2 \\
& & \\
\hline
& & \\
D_S^+ \rightarrow \pi^+ \pi^+ \pi^-  & 1.4 \pm 0.5  & 0.0003 \\
& & \\
\hline
& & \\
& & ^{a)} 0.7 \\
D_S^+ \rightarrow \pi^+ \pi^{\circ} \eta  & < 4.1  &  \\
& & ^{b)} 0.6 \\
& & \\
\hline
& & \\
& & ^{a)} 0.1 \\
D_S^+ \rightarrow \pi^+ \pi^{\circ} \eta' & < 4.2 &  \\
& & ^{b)} 0.1 \\
& & \\
\hline
& & \\
D_S^+ \rightarrow K^+ K^- \pi^+ & 1.2 \pm 0.5 & 0.2 \\
& & \\
\hline
\end{array}
$$
\newpage
\begin{center}
{\bf Table II}
\end{center}
$$
\begin{array}{|c|c|c|}
\hline
\hline
\multicolumn{1}{|c}{Decay} &
\multicolumn{1}{|c}{\Gamma_{exp} \times 10^{15} \; \, (GeV)} &
\multicolumn{1}{|c|}{\Gamma_{theor} \times 10^{15} \; \, (GeV)} \\
\hline
\hline
& & \\
D^{\circ} \rightarrow \pi^+ \pi^- \pi^{\circ} & -- & 1.7 \\
& & \\
\hline
& & \\
D^{\circ} \rightarrow K^+ K^- \pi^{\circ} & --  & 0.2 \\
& & \\
\hline
& & \\
D^{\circ} \rightarrow K^{\circ} K^- \pi^+ &   3.8 \pm 3.8 & 0.1 \\
& & \\
\hline
& & \\
D^{\circ} \rightarrow \overline{K}^{\circ} K^+ \pi^- &  6.3^{+3.8}_{-3.2}
  & 0.2 \\
& & \\
\hline
& & \\
D^+ \rightarrow \pi^+ \pi^+ \pi^-    & 1.6 \pm 0.4 & 1.2 \\
& & \\
\hline
& & \\
& & ^{a)} 0.1 \\
D^+ \rightarrow \pi^+ \eta \eta &   --  & \\
& & ^{b)} 0.2 \\
& & \\
\hline
& & \\
& & ^{a)} 0.2 \\
D^+ \rightarrow  \pi^+ \eta \eta' & --  &  \\
& & ^{b)} 0.2 \\
& & \\
\hline
& & \\
D^+ \rightarrow K^+ K^- \pi^+  & 2.9 \pm 0.6  & 0.1 \\
& & \\
\hline
& & \\
D_S^+ \rightarrow K^+ \pi^+ \pi^-  & -- & 0.3 \\
& & \\
\hline
\end{array}
$$
\newpage
\begin{center}
{\bf Table III}
\end{center}
$$
\begin{array}{|c|c|c|}
\hline
\hline
\multicolumn{1}{|c}{Decay} &
\multicolumn{1}{|c}{\Gamma_{exp} \times 10^{16} \; \, (GeV)} &
\multicolumn{1}{|c|}{\Gamma_{theor} \times 10^{16} \; \, (GeV)} \\
\hline
\hline
& & \\
D^{\circ}  \rightarrow K^{\circ} K^+ K^-  & -- & 0.02 \\
& & \\
\hline
& & \\
D^+ \rightarrow K^+ \pi^+ \pi^- & -- & 0.2 \\
& & \\
\hline
& & \\
D^+ \rightarrow K^+ K^+ K^-  \; ^{\cite{WA82}} &  18 \pm 9  &   0.001 \\
& & \\
\hline
\end{array}
$$

\newpage
\hspace{-0.6cm}{\Large \bf Figure Captions} \\
    \\
    \\
{\bf Figure 1}: Diagrams for the processes $D \rightarrow P P P$:
                a) external weak transition
                b) internal weak transition. $\Sigma$ is a scalar
                 meson and $\Pi$ is a pseudoscalar meson.
\\
\\
{\bf Figure 2}: a) Weak transition in the initial leg . Annihilation--like
                   diagram. $\Pi_1$ is a non--charmed pseudoscalar meson.
                b) Weak transition in a final leg. $\Pi_2$ is a charmed
                   pseudoscalar meson.

\newpage
\begin{figure}[h]
   \begin{Feynman}{230,60}{25,0}{0.7}
   \put(30,35){\fermionright}      \put(35,37){$D$}
   \put(60,35){\fermionright}  \put(75,37){$\Pi$}
   \put(91,35){\circle{2}}
   \put(92,35){\gaugebosonrighthalf}   \put(98,37){$s^{\pm}$}
   \put(108,35){\circle{2}}
   \put(109,35){\fermionrighthalf}  \put(115,37){$P_2$}
   \put(50,35){\fermiondrr}        \put(83,20){$P_3$}
   \put(80,50){\fermiondll}        \put(83,51){$P_1$}
   \put(140,35){\fermionright}     \put(145,37){$D$}
   \put(170,35){\fermionrighthalf}     \put(176,37){$\Sigma$}
   \put(185,35){\fermiondr}      \put(202,20){$P_3$}
   \put(185,35){\fermionrighthalf} \put(193,37){$\Pi$}
   \put(180,50){\fermionur}    \put(183,50){$P_1$}
   \put(201,35){\circle{2}}
   \put(202,35){\gaugebosonrighthalf}  \put(210,37){$s^{\pm}$}
   \put(218,35){\circle{2}}
   \put(219,35){\fermionrighthalf}   \put(225,37){$P_2$}
   \end{Feynman}

\end{figure}
\vspace*{-1cm}
\begin{center}
{\bf Figure 1.a)}
\end{center}
\vspace*{1cm}
\begin{figure}[h]
   \begin{Feynman}{225,60}{25,0}{0.7}
   \put(30,35){\fermionright}       \put(35,38){$D$}
   \put(68,50){\fermionur}          \put(72,50){$P_1$}
   \put(60,35){\fermionrighthalf}   \put(65,38){$\Sigma$}
   \put(76,35){\circle{2}}
   \put(77,35){\gaugebosonrighthalf}  \put(83,38){$s^{\pm}$}
   \put(93,35){\circle{2}}
   \put(94,35){\fermionrighthalf}   \put(99,38){$\Sigma$}
   \put(124,50){\fermionur}         \put(125,50){$P_2$}
   \put(109,35){\fermiondr}         \put(125,20){$P_3$}
   \put(145,35){\fermionright}      \put(150,38){$D$}
   \put(190,50){\fermionur}         \put(193,50){$P_1$}
   \put(175,35){\photonright}       \put(190,38){$W^{\pm}$}
   \put(205,35){\fermiondr}         \put(222,20){$P_3$}
   \put(220,50){\fermionur}         \put(222,50){$P_2$}
   \end{Feynman}
\end{figure}
\vspace*{-1cm}
\begin{center}
{\bf Figure 1.b)}
\end{center}
\newpage
\vspace*{1cm}
\begin{figure}[h]
   \begin{Feynman}{110,60}{25,0}{0.9}
   \put(40,35){\fermionright}       \put(45,38){$D$}
   \put(71,35){\circle{2}}
   \put(72,35){\gaugebosonrighthalf}  \put(78,38){$s^{\pm}$}
   \put(88,35){\circle{2}}
   \put(89,35){\fermionrighthalf}   \put(95,38){$\Pi_1$}
   \put(102,34){$\otimes$}
   \end{Feynman}
\end{figure}
\vspace*{-1cm}
\begin{center}
{\bf Figure 2.a)}
\end{center}
\vspace*{2cm}
\begin{figure}[h]
   \begin{Feynman}{110,60}{25,0}{0.9}
   \put(39,34){$\otimes$}
   \put(41,35){\fermionrighthalf}       \put(47,38){$\Pi_2$}
   \put(57,35){\circle{2}}
   \put(58,35){\gaugebosonrighthalf}  \put(65,38){$s^{\pm}$}
   \put(74,35){\circle{2}}
   \put(75,35){\fermionright}   \put(95,38){$P$}
   \end{Feynman}
\end{figure}
\vspace*{-1cm}
\begin{center}
{\bf Figure 2.b)}
\end{center}
\end{document}